\documentclass[prc,twocolumn,superscriptaddress,showpacs,showkeys,amsfonts,amssymb,amsmath]{revtex4}
\usepackage{graphicx}% Include figure files
\usepackage{dcolumn}% Align table columns on decimal point
\usepackage{bm}% bold math

\begin{document}

\title{\boldmath
Search for Narrow Nucleon Resonances below Pion Threshold in the
H(e,e$'\pi^{+}$)X and D(e,e$'$p)X Reactions
}

\author{M.~Kohl}
\altaffiliation[Present address:]{$\;$MIT-Bates Linear Accelerator,
 Middleton, \mbox{MA 01949}, U.S.A.}
\affiliation{Institut f\"{u}r Kernphysik, 
  Technische Universit\"at Darmstadt, D-64289 Darmstadt,
  Germany}
\author{M.~Ases~Antelo}
\author{C.~Ayerbe}
\author{D.~Baumann}
\author{R.~B\"ohm}
\affiliation{Institut f\"{u}r Kernphysik, 
  Universit\"{a}t Mainz,  D-55099 Mainz,
  Germany}
\author{D.~Bosnar}
\affiliation{Department of Physics,
  University of Zagreb, HR-10002 Zagreb, Croatia}                          
\author{M.~Ding}
\author{M.O.~Distler}
\author{J.~Friedrich}
\author{J.~Garc\'{\i}a~Llongo}
\author{P.~Jennewein}
\author{G.~Jover~Ma\~nas}
\author{H.~Merkel}
\author{P.~Merle}
\author{U.~M\"{u}ller}
\author{R.~Neuhausen}
\author{L.~Nungesser}
\author{R.~P\'erez~Benito}
\author{J.~Pochodzalla}
\affiliation{Institut f\"{u}r Kernphysik, 
  Universit\"{a}t Mainz,  D-55099 Mainz,
  Germany}
\author{M.~Potokar}
\affiliation{Institute ``Jo\v{z}ef Stefan'', 
  University of Ljubljana,
  SI-1001 Ljubljana, Slovenia}
\author{C.~Rangacharyulu}
\affiliation{Department of Physics, 
  University of Saskatchewan, Saskatoon,
  SK, S7N 5E2, Canada}
\author{A.~Richter}
\thanks{Corresponding author}
\email[$\;$email:$\;\;$]{richter@ikp.tu-darmstadt.de}
\affiliation{Institut f\"{u}r Kernphysik, 
  Technische Universit\"at Darmstadt, D-64289 Darmstadt,
  Germany}
\author{G.~Schrieder}
\affiliation{Institut f\"{u}r Kernphysik, 
  Technische Universit\"at Darmstadt, D-64289 Darmstadt,
  Germany}
\author{M.~Seimetz}
\author{Th.~Walcher}
\author{M.~Weis}
\affiliation{Institut f\"{u}r Kernphysik, 
  Universit\"{a}t Mainz,  D-55099 Mainz,
  Germany}

\date{\today}

\begin{abstract}
% Text of abstract
In two series of high-resolution coincidence experiments at the
three-spectrometer facility at MAMI, 
the H(e,e$'\pi^{+}$)X and D(e,e$'$p)X
reactions were studied to search for narrow
nucleon resonances below pion threshold. The missing-mass 
resolution was 0.6 to 1.6 MeV/c$^{2}$ (FWHM) in the proton experiment and
0.9 to 1.3 MeV/c$^{2}$ in the deuteron experiment.
The experiments covered the missing-mass region from the neutron
mass up to about 1050 and 1100 MeV/c$^{2}$, respectively. 
None of our measurements showed a signal for narrow resonances to a level of
down to 10$^{-4}$ 
with respect to the neutron peak in the missing-mass spectra.
\end{abstract}



\maketitle

% main text
\section{Introduction}
\label{sec:introduction}
It is an important issue of modern experiments in hadron physics
to provide significant information on manifestations of QCD. 
As is well known (see e.g.~\cite{thomasweise}),
the description of the experimentally observed
baryon excitation spectrum in terms of the 
colorless three-quark ($q^{3}$) configuration within
the various quark models has been quite successful.
Moreover, QCD allows for exotic structures like
multi-quark configurations, glueballs and hybrids. 
Such states are mainly
expected in the mass region above 1.5 GeV/c$^{2}$. 
It is one of the irritating facts, however,  
that already the $q^{3}$ description
predicts far more resonance states than experimentally observed.
Thus, the search for missing resonances is one of the
major concerns of hadron experiments. 
On the other hand, the existing
low-lying states (nucleon, $\Delta$, N$^{*}$) are well accounted for within
the three-quark picture and 
neither QCD-inspired calculations at low momentum and 
close to the chiral limit (i.e. Chiral Perturbation Theory) nor perturbative
QCD in the high-momentum regime predict states below or close to
pion threshold.
If such low-lying states did exist, configurations beyond the naive $q^{3}$
picture would be required for an explanation. 

Since the strong decay via pion emission is energetically forbidden
for states below pion threshold, these states would
necessarily be narrow (less than a few keV) 
with lifetimes typical for electromagnetic or weak decays. 
A precise examination of the low-lying nucleon excitation spectrum with
various probes is thus 
a powerful tool to test QCD.

Recently, there have been two experiments reported in the literature,
one performed at the
SATURNE synchrotron in Saclay and  the other at the INR in Moscow, which
claimed the observation of excited states of the nucleon below pion
threshold~\cite{tati,filkov}. 
Tati\-scheff {\it et al.} utilized the H(p,p$'\pi^{+}$)X reaction to
measure the invariant-mass spectrum of the missing nucleon~\cite{tati}. 
They claimed to
have found resonance states at $M_{X}=1004$, 1044, and 1094 MeV/c$^{2}$, with
a statistical significance of up to 17 standard deviations. 
However, the missing-mass resolution of this experiment was
limited and ranged from 7 to 25 MeV/c$^{2}$ (FWHM). Very recently, Tatischeff 
{\it et al.} 
reported further
the observation of additional states at 1136, 1173, 1249, 1277, 1384,
and possibly 1339 MeV/c$^{2}$  
using the H(p,p$'\pi^{+}$)X, the D(p,p$'$p)X, and the H(p,p$'$p)
reactions~\cite{tati2}.

Fil'kov {\it et al.}~\cite{filkov} 
found resonance structures in the D(p,p$'$p)X
reaction in the spectra of
the invariant masses $M_{pX}$ as well as $M_{X}$ and claimed the observation of
``supernarrow'' dibaryons (SND) at $M_{pX} = 1904$, 1926, and 1942
MeV/c$^{2}$.
The additionally observed resonance structures at $M_{X} = 966$, 986, and
1003 MeV/c$^{2}$ differ from those reported in $M_{pX}$ by about 
one nucleon mass. 
They were interpreted to be either genuine, i.e. SND $\rightarrow$ p + X, 
or to result from the decay SND $\rightarrow$ p + n + $\gamma$ under the
restrictions of a small phasespace and experimental acceptance.
The significance for the SND peaks was estimated to be  6~to~7 standard
deviations, with a moderate missing-mass resolution of about 5 MeV/c$^{2}$.

So far, various attempts for an explanation of the narrow
resonances~\cite{tati,filkov} 
have been offered. Model assumptions beyond the $q^{3}$ picture result
in mass formula prescriptions based on
colored~\cite{mulders} and colorless~\cite{konno} quark clusters.
It has been argued however~\cite{lvov}, that 
narrow nucleon resonances should be completely excluded, since they were not
observed in real Compton scattering experiments~\cite{leon}.
Another approach describes the narrow resonances as
possible members of the SU(6) spin-flavor multiplet, where only
the two-photon transition channel is allowed~\cite{kobushkin}. 
This picture accounts for the fact that narrow resonances were not
observed in photon-induced reactions like in real
Compton scattering experiments. 
Recently, it was proposed to interpret
the narrow states as collective excitations of the
quark condensate through the multiple production of ``genuine''
Goldstone pions with a light mass of $m \approx$ 20
MeV/c$^{2}$~\cite{walcher}.  
In this model, the narrow states are spatially extended structures and
thus they are likely inaccessible to the real photon probe.

Another theoretical interest in the narrow baryons is their influence on
astrophysical objects. It was pointed out that the existence of 
narrow nucleon resonances would drastically affect the
structure of neutron stars by lowering their possible maximal mass,
which would be inconsistent with the observations~\cite{kolomeitsev}.  
However, the existence of light pions
would still be compatible with the observed masses of neutron stars. 

Meanwhile, there have been reported experiments in direct response to
Tatischeff's and Fil'kov's findings.
In a pilot study of the H(e,e$'\pi^{+}$)X reaction at the Jefferson
Laboratory~\cite{jlab}, the yield ratio H(e,e$'\pi^{+}$)X /
H(e,e$'\pi^{+}$)n was measured, which excludes narrow nucleon resonances to a
level of 10$^{-3}$.
The D(p,p$'$p)X reaction has been revisited at the RCNP in Osaka~\cite{tamii},
yielding a null result for the above claimed states.
However, the significance of the result of the latter experiment
has been questioned in Ref.~\cite{filkov2}. 
In fact, we have also reanalyzed the data of the virtual Compton scattering 
experiments at MAMI, but we did not find any signal above a significance of
three standard deviations.
In a photoproduction experiment at MAMI, the reactions
\mbox{H($\gamma$, $\pi^{+}\gamma$n)}
and H($\gamma$, $\pi^{+}\gamma\gamma$n) 
have been measured with a high luminosity and results are awaited~\cite{beck}. 

The issue of the occurence of resonances below pion threshold is thus
unsettled. 
The observation of such narrow states is always limited by the
achievable resolution. Experiments providing highest resolution are
therefore best suited to perform such studies. 
In this paper, we report the results of two series of high-resolution
measurements using the 
H(e,e$'\pi^{+}$)X and D(e,e$'$p)X reactions. 
The first reaction has been chosen as the analogous reaction to
the 
H(p,p$'\pi^{+}$)X
process of Tati\-scheff {\it et al.}~\cite{tati}.  
The energy transfer corresponds to the excitation region of the $\Delta$
resonance. 
With the detection of the scattered electron in coincidence with the produced
pion the missing mass of the reaction has been
reconstructed. The experiment covered the missing-mass region from the neutron
mass up to about 1050 MeV/c$^{2}$. 
In addition, 
the second electromagnetic reaction D(e,e$'$p)X has been studied
as the analog of the strong D(p,p$'$p)X
experiment by Fil'kov {\it et al.}~\cite{filkov}.
The energy transfer varied from 
close to the threshold of the deuteron breakup up to the quasielastic region. 
With the detection of both the scattered electron and the emitted proton,
the missing mass of the reaction has been reconstructed. 
The experiment covered the missing-mass region from the neutron
mass up to about 1100 MeV/c$^{2}$.

\section{Measurements}
\label{sec:experiment}
The measurements were carried out at the three-spectrometer facility of the A1
collaboration at the Mainz Microtron MAMI~\cite{MAMI}.
The target was a standard liquid hydrogen/deuterium target, which was
operated at beam currents of up to 15~$\mu$A. The target
cell has been
made of a thin HAVAR foil of 10.2 $\mu$m thickness (8.9 $\mu$m in case
of the deuterium target). The cell geometry is stadion-like  with a length
of 5 cm along the beam axis and a width of 1 cm in the perpendicular direction.
The beam was swept before entering the scattering chamber to fill a square of
4$\times$4 mm$^{2}$ on the target. 
The three magnetic spectrometers A, B, and C have been
used for the detection of the scattered electrons and of
the pions or protons of the respective reactions. The spectrometers are
equipped with vertical drift chambers for tracking, a thin and a thick layer
of plastic scintillators for the trigger and for 
$\pi^{+}$/p discrimination, and a large-volume gas
\v Cerenkov counter for e$^{\pm}/\pi^{\pm}$ identification. The setup allows 
for very clean and high-resolution ($\Delta p/p \lesssim 10^{-4}$,
$\Delta \theta \lesssim 3$ mrad) coincidence measurements
yielding missing-mass resolutions of typically 1 MeV/c$^{2}$ (FWHM).

\subsection{The H(e,e$'\pi^{+}$)X Reaction}
\label{sec:heepx}

The first pilot measurement in the H(e,e$'\pi^{+}$)X reaction based on our
proposal to the Program Advisory Committee 
at MAMI~\cite{prop98} was carried out
in July 2000 with a total runtime of
18 h. The missing-mass spectrum showed a
resonance-like structure around $M_{X} \approx 1007$ MeV/c$^{2}$, with a width
of about 1 MeV/c$^{2}$, i.e. with the resolution of the setup. However,
the significance was only $\approx 3$ standard deviations, not warranting 
an interpretation of this finding as a positive identification.

A dedicated
H(e,e$'\pi^{+}$)X experiment was thus 
performed in February and August 2002 with a
total runtime of 257 hours.
The parameters of the 
kinematical settings are listed 
in Tab.~\ref{tab:protonkins}.  
\squeezetable
\begin{table}[b]
\caption{Kinematics of the H(e,e$'\pi^+$)X reaction. 
  The columns
  $p_{e'}$ ($\theta_{e'}$) and $p_{\pi^{+}}$ ($\theta_{\pi^{+}}$) are the
  central momenta and angles of the spectrometers. Also shown are
  the four-momentum transfer $Q^{2}$, the $\gamma^{*}$p invariant mass $W$, 
  and the pion center-of-mass angle
  $\theta_{\pi}^{*}$. The average beam current $I$ and the net runtime $T$ are
  shown in the rightmost columns. The settings {\it H-1, 2,} and {\it 3}
  refer to the measurements with Spectrometer A and B; settings  {\it H-4, 5,}
  and {\it 6} with B and C, respectively.
}
 \label{tab:protonkins}
\setlength{\tabcolsep}{0.06cm}
 \begin{tabular}{l|ccccc|ccccr}
\hline\hline
Kin   & $E_{0}$ & $\theta_{e'}$ &  $\theta_{\pi^{+}}$ & $p_{e'}$ 
& $p_{\pi^{+}}$ & $Q^{2}$                     & $W$            
&  $\theta_{\pi}^{*}$ &  $I$     & $T$\\
          & [MeV]   &               &             & [$\frac{\rm MeV}{\rm c}$]  
& [$\frac{\rm MeV}{\rm c}$] & [$\frac{\rm GeV^2}{\rm c^2}$] 
& [$\frac{\rm MeV}{\rm c^2}$]
&                     & [$\mu$A] & [h]\\
\hline
(AB) & & & & &\\
% NB_1_AB
{\it H-1} &855.1 & 17.4$^{\circ}$ & 22.4$^{\circ}$ & 510 & 290
& 0.04 & 1220 & 0$^{\circ}$ & 5.3 & 2.0\\
% NB-2
%H-2 &855.1 & 17.4$^{\circ}$ & 22.4$^{\circ}$ & 510 & 225
%& 0.04 & 1220 & 0$^{\circ}$ & 5.7 & 17.2\\
% NB_2_AB
{\it H-2} &855.1 & 17.4$^{\circ}$ & 22.4$^{\circ}$ & 510 & 225
& 0.04 & 1220 & 0$^{\circ}$ & 7.0 & 50.8\\
% NB_3_AB
{\it H-3} &855.1 & 17.4$^{\circ}$ & 22.4$^{\circ}$ & 510 & 236
& 0.04 & 1220 & 0$^{\circ}$ & 6.7 & 26.4\\
\hline
(BC) & & & & &\\
% NB_1_BC
{\it H-4} &855.1 & 17.4$^{\circ}$ & 55.0$^{\circ}$ & 510 & 215
& 0.04 & 1220 & 100$^{\circ}$ & 5.3 & 2.2\\
% NB_2_BC
{\it H-5} &855.1 & 17.4$^{\circ}$ & 55.0$^{\circ}$ & 510 & 150
& 0.04 & 1220 & 100$^{\circ}$ & 6.9 & 79.8\\
% NB_3_BC
{\it H-6} &659.7 & 20.3$^{\circ}$ & 88.4$^{\circ}$ & 150 & 290
& 0.0125 & 1350 & 110$^{\circ}$ & 14.3 & 96.0\\
\hline\hline
 \end{tabular}
 \end{table}
The scattered electrons were detected in Spectrometer
B at forward angles, and both Spectrometer A and C were used in coincidence
with B to detect the pions along the momentum-transfer direction
(pion center-of-mass angle $\theta_{\pi}^{*} \approx 0^\circ$), and almost
perpendicular to it 
($\theta_{\pi}^{*} \approx 100^\circ$ and $110^{\circ}$), respectively. 
Most of the AB and BC measurements were carried out simultaneously,
exploiting the full capabilities of the three-spectrometer facility.

The resolution of the measured coincidence time,
%(i.e. the time difference
%between the detection of the two particles belonging to the reaction) 
after
applying corrections due to individual path lengths and detector-related
influences on the timing signals, varied from 0.8 to 1.2 ns (FWHM). 
The true events were selected by a $\pm 1.5$ ns 
cut on the coincidence
time spectrum. The shape of the background has been determined by selecting
purely random coincidences. 
%An example spectrum of the missing mass under true and random cut condition 
%and after applying particle identification cuts
%is shown in Fig.~\ref{fig:raw-NB-3-BC} a).
Furthermore, the data were normalized to the collected charge and the target
thickness  and were corrected for deadtime effects. The loss of pions due to
their decay in flight from the origin to the detector has been corrected
by weighting each pionic event with an individual survival probability,
depending on the pion momentum and on the length of the pion track.
The acceptance was determined by a simulation code, which accounts
for the energy loss in the target and window materials of the scattering
chamber and the spectrometer entrance as well as for the Bethe-Heitler
radiation of the electrons. 
%The event generator started from uniform
%distributions of the detected angles and momenta in the two spectrometers. 
%The shape of the missing-mass acceptance is shown in
%Fig.~\ref{fig:raw-NB-3-BC} b).
After subtraction of the random background  and application of the mentioned
corrections, normalized yields for the H(e,e$'\pi^{+}$)X reaction 
have been obtained.

\begin{figure}[t]
\centering\includegraphics[angle=90,scale=1]{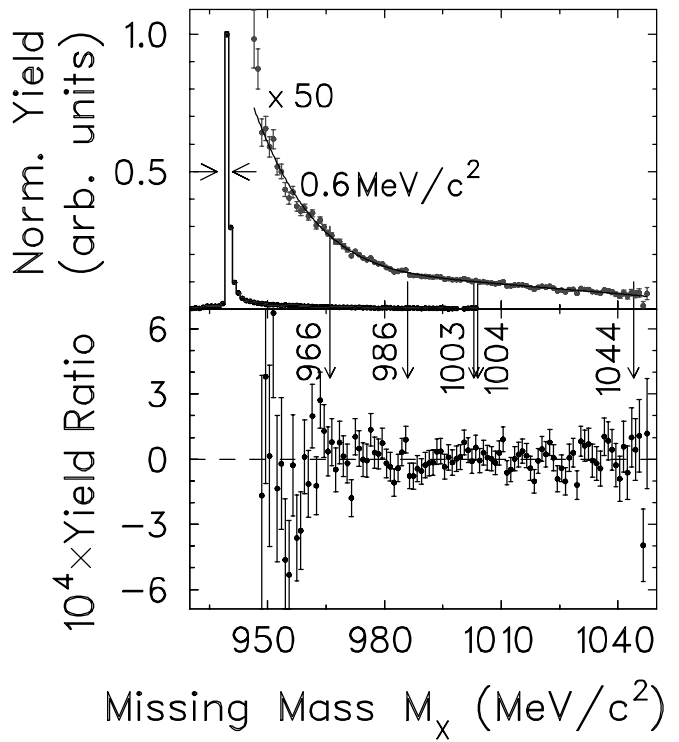}
\caption{Upper half: Missing-mass spectrum
  of the H(e,e$'\pi^+$)X reaction, % after subtracting the
%  random background and applying the corrections mentioned in the text,
  measured with Spectrometer A and B at $\theta_{\pi}^{*} = 0^{\circ}$.
  The plot shows the result of setting {\it H-1} (with the neutron peak height
  set to unity) and the combined settings {\it H-2} and {\it H-3} (at
  the electron beam energy and spectrometer settings listed in
  Tab.~\ref{tab:protonkins}). The achieved missing-mass resolution is 0.6
  MeV/c$^{2}$ (FWHM) as given by the width of the neutron peak.
  A fifth-order polynomial is fit to the data to model the radiation tail. 
  The invariant masses of the states claimed by Refs.~\cite{tati,filkov} are
  indicated by the arrows. 
  Lower half: Yield ratio of the above spectrum after subtraction of
  the radiation
  tail and division by the height of the neutron peak.
  Narrow structures (see the experiments of~\cite{tati} and~\cite{filkov})
  are excluded to a level of 10$^{-4}$
  of the height of the neutron peak.
}
\label{fig:heepix1}
\end{figure}

\begin{figure*}
\centering\includegraphics[angle=90,scale=1]{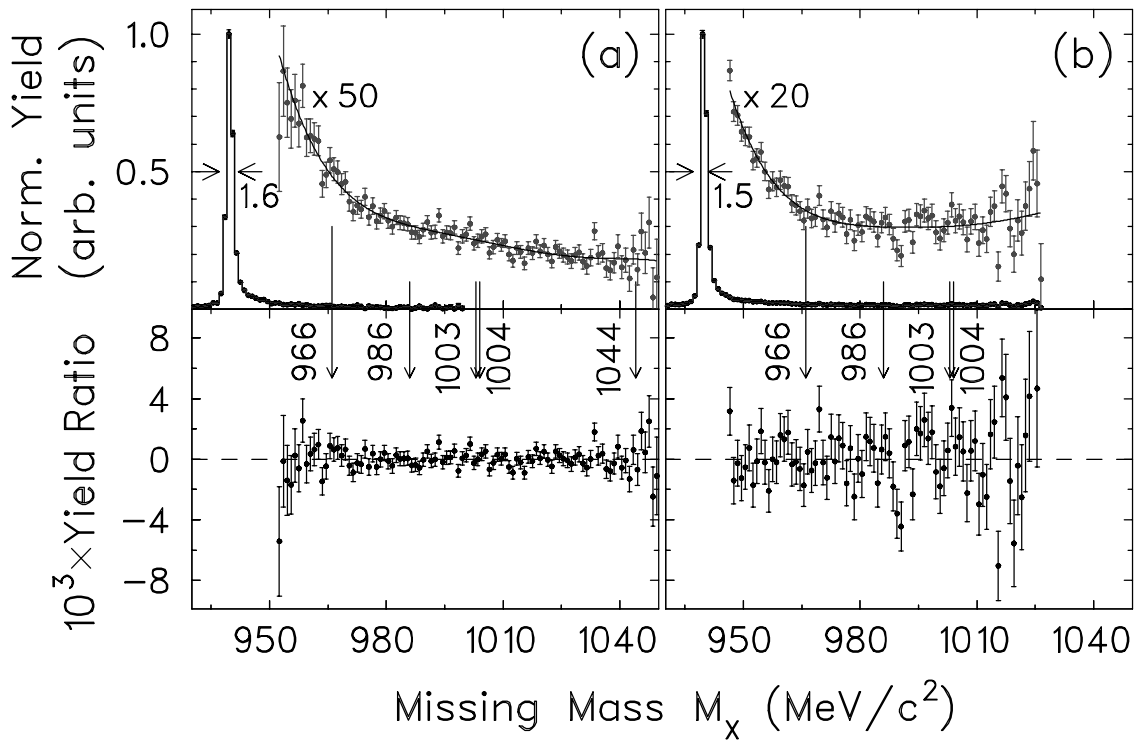}
\caption{Upper half: Missing-mass spectra
  of the H(e,e$'\pi^+$)X reaction,
  measured with Spectrometer B and C at $\theta_{\pi}^{*} \approx
  100^{\circ}$.  
  Plot (a) shows the settings {\it H-4} (with the neutron peak height
  set to unity) and {\it H-5}, and  (b) setting {\it  H-6} (at 
  the electron beam energy and spectrometer settings listed in
  Tab.~\ref{tab:protonkins}). The achieved missing-mass resolution is (a)
  $\approx$ 1.6 MeV/c$^{2}$ (FWHM) 
  and (b) $\approx$ 1.5 MeV/c$^{2}$, respectively.
  The vertical arrows are the same as in Fig.~\ref{fig:heepix1}. The solid
  curves and the yield ratios in the lower half are obtained in the same way
  as in  Fig.~\ref{fig:heepix1}. 
  Narrow structures (see the experiments of~\cite{tati} and~\cite{filkov})
  are excluded to a level of (a) 10$^{-3}$ and (b) $2
  \cdot 10^{-3}$ of the height of the neutron peak, respectively.
}
\label{fig:heepix2}
\end{figure*}

\begin{figure*}
\centering\includegraphics[angle=90,scale=1]{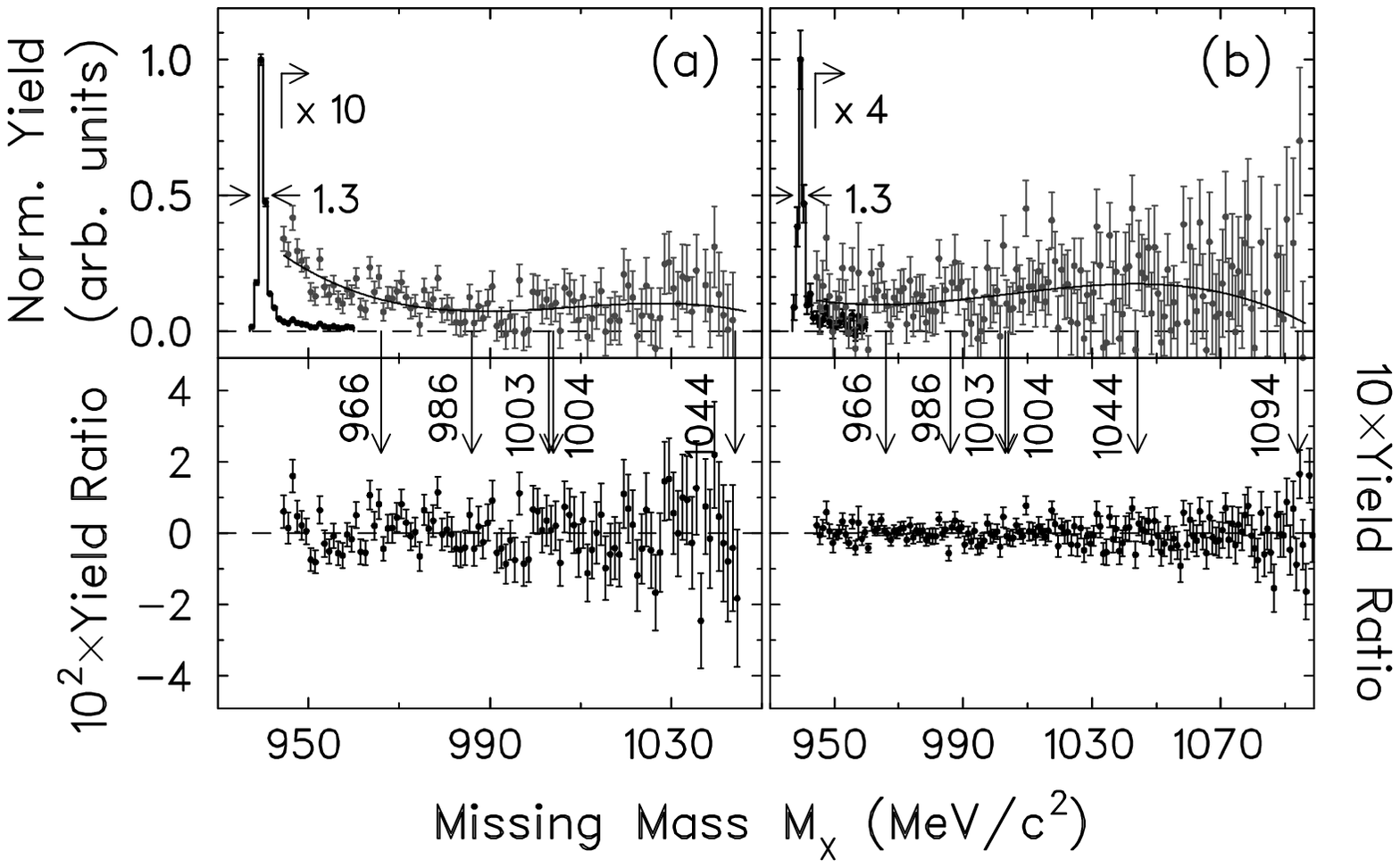}
\caption{Upper half: Missing-mass spectra
  of the D(e,e$'$p)X reaction in the region of the deuteron-breakup threshold
  at the four-momentum transfer (a) $Q^{2} = 0.8$ (GeV/c)$^{2}$ 
  and (b) $Q^{2} = 1.1$ to 1.2 (GeV/c)$^{2}$
  (with the neutron peak height set to unity).
  The plot shows the combined settings (a) {\it D-1,2,3}, and (b) {\it D-6,7}
  (cf. Tab.~\ref{tab:deuteronkins}).
  The achieved missing-mass resolution is $\approx$ 1.3 MeV/c$^{2}$. The solid
  curves and the yield ratios in the lower half are obtained  in the same way
  as in Fig.~\ref{fig:heepix1}. 
  Significant structures are excluded to a level of (a)
  10$^{-2}$ and (b) $5 \cdot 10^{-2}$ of the height of the
  neutron peak, respectively.
}
\label{fig:deepx1}
\end{figure*}
Figures~\ref{fig:heepix1} and~\ref{fig:heepix2} summarize 
the results of the H(e,e$'\pi^{+}$)X
measurements. In the upper halves of Figs.~\ref{fig:heepix1}
and~\ref{fig:heepix2}, the normalized yields
of three kinematical settings are presented as a function of the missing mass. 
The missing-mass resolution was $\approx$ 
0.6 MeV/c$^{2}$ (FWHM) for the AB runs shown in Fig.~\ref{fig:heepix1}, 
and $\approx$ 1.5 to 1.6 MeV/c$^{2}$ for the BC runs plotted in 
Fig.~\ref{fig:heepix2}.
The invariant masses of the states claimed by 
Tatischeff {\it et al.}~\cite{tati} and Fil'kov {\it et al.}~\cite{filkov} are
indicated by the vertical arrows. 
As is seen from the spectra, no significant narrow structure other than the 
peak corresponding to the ground state of the neutron is found.
The missing-mass region from $M_{X} \approx 950$ to 1050 MeV/c$^{2}$ 
shows a smooth radiation tail. 
The measurements include a repetition of the
kinematics of the pilot run (settings {\it H-2/H-3} in Fig.~\ref{fig:heepix1})
with about five times the statistics of the first run. 
In order to quantify the sensitivity of our measurements, we
show yield ratios in the lower halves of Figs.~\ref{fig:heepix1}
and~\ref{fig:heepix2}.  They are obtained by subtracting 
the radiation tail, modelled by a fifth-order polynomial,
and dividing the spectra by the height of the respective
neutron peak. 
The differential cross section for the H(e,e$'\pi^{+}$)n channel 
can be estimated using the Unitary Isobar Model ``MAID''~\cite{maid}. 
For the three kinematics
of Figs.~\ref{fig:heepix1}, \ref{fig:heepix2}(a), and \ref{fig:heepix2}(b), 
one obtains
$\rm d^{3}\sigma / (d\Omega_{e'}dE_{e'}d\Omega_{\pi}^{*}) \approx $
$2.6 \cdot 10^{-4},\, 8.2 \cdot 10^{-4}$, and 
$1.9 \cdot 10^{-5}$ $\mu \rm b / (\rm{sr}^{2} \rm{MeV})$. 
As the yield ratios in the lower halves 
indicate, narrow structures can be excluded to a level of 10$^{-4}$,
10$^{-3}$, and 2$\cdot 10^{-3}$ 
with respect to the elementary pion production process leaving the 
neutron in the ground state, respectively.

\begin{figure*}
\centering\includegraphics[angle=90,scale=1]{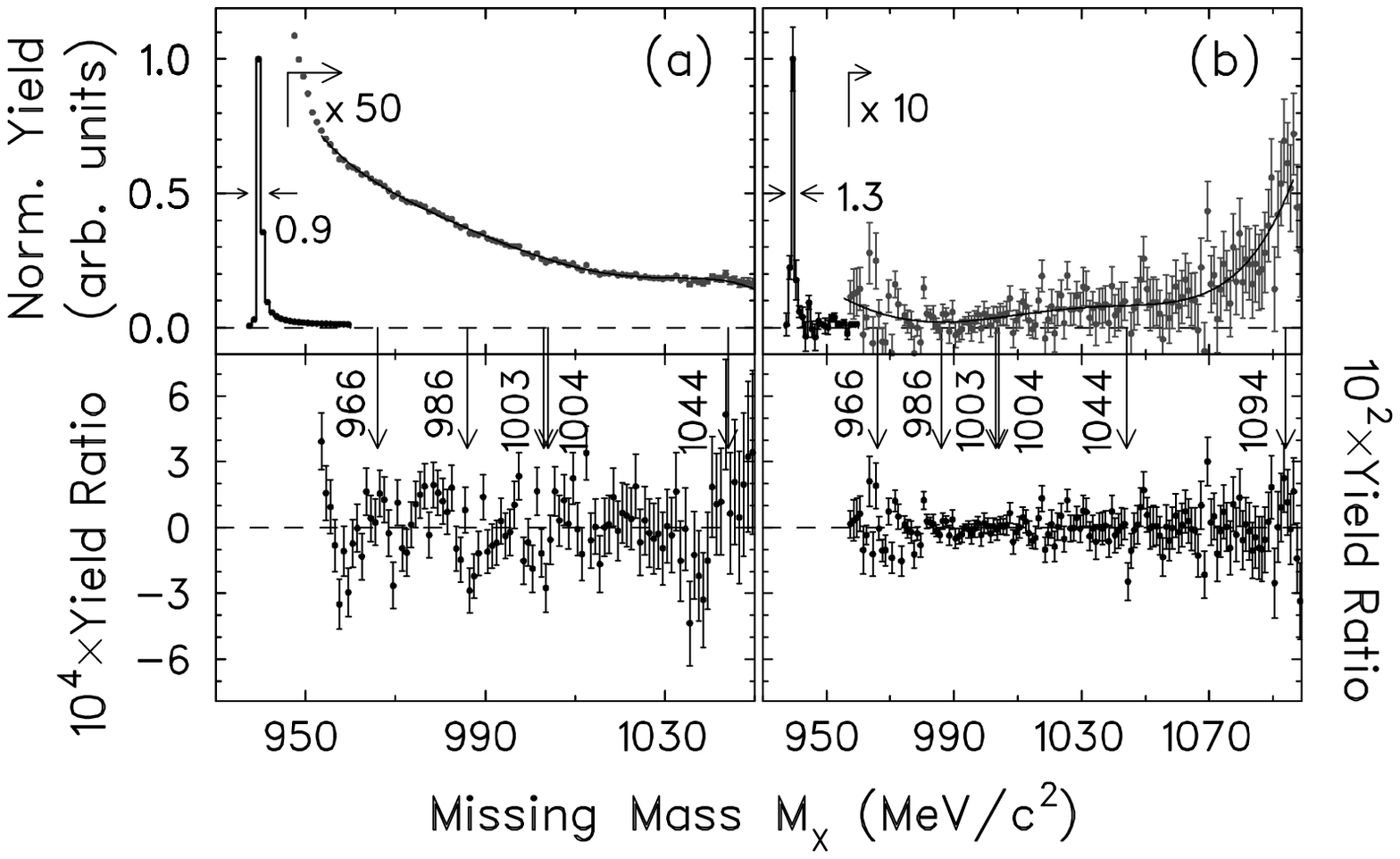}
\caption{Upper half: Missing-mass spectra
  of the D(e,e$'$p)X reaction in the region near the quasielastic peak
  at (a) $Q^{2} = 0.6$ to 0.7 (GeV/c)$^{2}$ 
  and (b) $Q^{2} = 0.8$ to 1.1 (GeV/c)$^{2}$
  (with the neutron peak height set to unity).
  The plot shows the combined settings (a) {\it D-4,5}, and (b) {\it D-8,9,10}
  (with the spectrometer settings listed in Tab.~\ref{tab:deuteronkins}).
  The achieved missing-mass resolution is (a) $\approx$ 0.9 MeV/c$^{2}$ and
  (b) $\approx$ 1.3 MeV/c$^{2}$. The solid
  curves and the yield ratios in the lower half are obtained in the same way
  as in
  Fig.~\ref{fig:heepix1}. Significant structures are excluded to a level of (a)
  10$^{-4}$ and (b) 10$^{-2}$ of the height of the
  neutron peak, respectively.
}
\label{fig:deepx2}
\end{figure*}

\subsection{\boldmath The D(e,e$'$p)X Reaction} 
\label{sec:deepx}

The D(e,e$'$p)X reaction was studied between May and August 2002 with a total
runtime of 632 hours.
The measured kinematics can be characterized
by the relative energy $E_{pX} = W - M_{p} - M_{X}$ of the p-X system,
where $W$ is the photon-deuteron invariant mass, $M_{p}$ the proton mass,
and $M_{X}$ the missing mass of the D(e,e$'$p)X reaction.
The kinematics of the measured settings varied from the threshold region
($E_{pX} \rightarrow 0$) up to the quasielastic region ($E_{pX} \approx 100$
to 150 MeV).  
The near-threshold kinematics have been chosen 
in analogy to the experiment by Fil'kov
{\it et al.}~\cite{filkov}, 
where the relative energy of the two outgoing nucleons pn 
(or possible pX) is small and a strong final-state interaction is expected.
In this case, the detected proton is emitted into a small forward cone around
the direction of momentum transfer.
The kinematical parameters of the 
measured settings of the D(e,e$'$p)X reaction are listed 
in Tab.~\ref{tab:deuteronkins}.
\squeezetable
\begin{table}[b]
\caption{Kinematics of the D(e,e$'$p)X reaction.
  The columns
  $p_{e'}$ ($\theta_{e'}$) and $p_{p}$ ($\theta_{p}$) are the
  central momenta and angles of the electron spectrometer (C) and the proton
  spectrometer (B), respectively. Also shown are
  the four-momentum transfer $Q^{2}$ and the p-X relative energy $E_{pX}$. 
  The average beam current $I$ and the net runtime $T$ are
  shown in the rightmost columns. 
}
 \label{tab:deuteronkins}
\setlength{\tabcolsep}{0.09cm}
 \begin{tabular}{l|ccccc|cccr}
\hline\hline
Kin   & $E_{0}$ & $\theta_{e'}$ &  $\theta_{p}$ & $p_{e'}$ & $p_{p}$ &
          $Q^{2}$                     & $E_{pX}$ & 
          $I$      & $T$\\
          & [MeV]   &               &         & [$\frac{\rm MeV}{\rm c}$] 
& [$\frac{\rm MeV}{\rm c}$] &
          [$\frac{\rm GeV^2}{\rm c^2}$] & [MeV]    &
          [$\mu$A] & [h]\\
\hline
% 660-1a
%{\it D--1a} & 659.7 & 110$^{\circ}$ & 26$^{\circ}$ & 420 & 444
%& 0.8 & 0-6 & 15 & 49.5\\
% 660-1b
%{\it D--1b} & 659.7 & 110$^{\circ}$ & 26$^{\circ}$ & 420 & 444
%& 0.8 & 0-6 & 15 & 131.7\\
% 660-1a+1b
{\it D-1} & 659.7 & 110$^{\circ}$ & 26$^{\circ}$ & 420 & 444
& 0.8 & 0-6 & 15 & 181.2\\
% 660-2
{\it D-2}  & 659.7 & 110$^{\circ}$ & 26$^{\circ}$ & 420 & 504
& 0.8 & 0-18 & 15 & 60.2\\
% 660-3
{\it D-3}  & 659.7 & 110$^{\circ}$ & 26$^{\circ}$ & 400 & 580
& 0.7 & 10-45 & 12.5 & 68.5\\
% 660-4a
%{\it D--4a} & 659.7 & 110$^{\circ}$ & 26$^{\circ}$ & 400 & 665
%& 0.7 & 30-80 & 15 & 3.5\\
% 660-4b
%{\it D--4b} & 659.7 & 110$^{\circ}$ & 26$^{\circ}$ & 385 & 665
%& 0.7 & 25-90 & 15 & 11.2\\
% 660-4a+4b
{\it D-4}  & 659.7 & 110$^{\circ}$ & 26$^{\circ}$ & 385 & 665
& 0.7 & 25-90 & 15 & 14.7\\
% 660-5
{\it D-5}  & 659.7 & 103$^{\circ}$ & 26$^{\circ}$ & 355 & 740
& 0.6 & 75-145 & 12.5 & 29.5\\
\hline
% 883-8a
%{\it D-6}  & 883.1 &  107$^{\circ}$  & 25$^{\circ}$ & 485 & 570
%& 1.2 & 0-14 & 10 & 31.5\\
% 883-8b
%{\it D-6}  & 883.1 &  107$^{\circ}$ & 25$^{\circ}$  & 485 & 570
%& 1.2 & 0-14 & 10 & 181.0\\
%883-8a+8b
{\it D-6}  & 883.1 &  107$^{\circ}$ & 25$^{\circ}$  & 485 & 570
& 1.2 & 0-14 & 10 & 212.5\\
% 883-7
{\it D-7}  & 883.1 &  107$^{\circ}$  & 25$^{\circ}$ & 485 & 610
& 1.1 & 0-24 & 10 & 47.0\\
% 883-6
{\it D-8}  & 883.1 &  107$^{\circ}$  & 25$^{\circ}$ & 485 & 713
& 1.1 & 10-60 & 10 & 3.7\\
% 883-3
{\it D-9}  & 883.1 & 92.3$^{\circ}$  & 30$^{\circ}$ & 485 & 705
& 0.8 & 20-80 & 10 & 13.5\\
% 883-2
{\it D-10} & 883.1 & 92.3$^{\circ}$  & 30$^{\circ}$ & 485 & 805
& 0.8 & 60-130 & 10 & 1.0\\
\hline\hline
 \end{tabular}
 \end{table}

The D(e,e$'$p)X measurements were carried out simultaneously
with a dedicated experiment to measure 
the electric form factor
of the neutron $G_{E,n}$ via the D($\vec{\rm e}$,e$'\vec{\rm n}$)
reaction~\cite{gen}.
The $G_{E,n}$ experiment utilized 
Spectrometer A for the scattered electrons
and a separate setup to measure the neutron polarization.
Spectrometer C (at angles $\gtrsim 90^{\circ}$ right of the beam) 
and Spectrometer B (at forward angles left of the beam) 
were used in coincidence for the detection of electrons  and
protons, respectively, in a parasitic mode. 
The electron scattering angle was restricted to large
values, which, though not optimal, resulted in quite sizeable
four-momentum transfer $Q^{2}$. 
%$^{2} = (E_{0} - E_{e'} + m_{d} - E_{p})^{2} - (\vec{p}_{0} - \vec{p}_{e'}
%- \vec{p}_{p})^{2}$ the squared missing mass of the D(e,e$'$p)X reaction. 
%The quantities $E_{i} (\vec{p}_{i})$ denote the four-vector components of the
%incoming and outgoing electron, and the proton, respectively, and $m_{d}$
%is the deuteron rest mass.

The analysis of the missing-mass spectra has been carried out 
in a way very similar
to that of the H(e,e$'\pi^{+}$)X measurements described above.
To a very large extent, 
the true $\pi^{-}$p coincidence events, which occurred in
addition to the random background events, could be identified and removed 
by requiring a signal in the \v Cerenkov detector in the electron arm
(Spectrometer C).
%, as it is expected to give no signal for $\pi^{-}$ events. 
Most of the remaining true 
$\pi^{-}$p events could be identified
in the coincidence time spectrum, where the e$'$p and $\pi^{-}$p peaks are
separated by about 2 ns.

The resulting normalized yields and yield ratios of the D(e,e$'$p)X reaction 
are presented in Figs.~\ref{fig:deepx1} and~\ref{fig:deepx2}. 
Figure~\ref{fig:deepx1} shows
the results of the near-threshold settings, while the results of the
quasielastic 
settings are depicted in Fig.~\ref{fig:deepx2}.
The missing-mass resolution was $\approx$ 
1.3 MeV/c$^{2}$ (FWHM) for the near-threshold runs, 
and $\approx$ 0.9 (1.3) MeV/c$^{2}$ for the runs near the quasielastic maximum.
The continuous part of the
missing-mass distribution is mainly due to the radiation tail,
with the above mentioned $\pi^{-}$p background giving a small
contribution in the higher missing-mass region.
In Fig.~\ref{fig:deepx2}(b), the pion threshold is open above 
$M_{X} > 1070$ MeV/c$^{2}$, yielding additional strength.
The invariant masses of the states claimed by 
Tatischeff {\it et al.}~\cite{tati} and Fil'kov {\it et al.}~\cite{filkov} are
indicated by the vertical arrows. 
Here again, as in the case of the hydrogen experiment described above, no
significant peak structure other than the one corresponding to the ground
state of the neutron is found.
Though the measurements in the threshold region took most of the beamtime
(see Tab.~\ref{tab:deuteronkins}), the statistics is quite poor even for the
ground-state breakup of the deuteron.
This is due to the fact that this kinematics maps out very high missing
momenta of the proton in the order of 450 MeV/c. 
In the quasielastic settings (Fig.~\ref{fig:deepx2}), 
the cross section is much larger.
The differential cross section for the D(e,e$'$p)n channel 
can be estimated using the deuteron electrodisintegration model of
Arenh\"ovel {\it et al.}~\cite{arenhoevel}.
For the threshold settings of Fig.~\ref{fig:deepx1}(a)
and~\ref{fig:deepx1}(b), 
one obtains an average of
$\rm d^{3}\sigma / (d\Omega_{e'}dE_{e'}d\Omega_{p}^{*}) \approx $
$1.4 \cdot 10^{-6}\, (2.8 \cdot 10^{-7})$ 
$\mu \rm b / (\rm{sr}^{2} \rm{MeV})$, respectively,
whereas the average differential cross
section of the quasielastic settings of Fig.~\ref{fig:deepx2}(a)
and~\ref{fig:deepx2}(b) is about
$\rm d^{3}\sigma / (d\Omega_{e'}dE_{e'}d\Omega_{p}^{*}) \approx $
$1.1 \cdot 10^{-3}\, (2.3 \cdot 10^{-5})$ $\mu \rm b / (\rm{sr}^{2} \rm{MeV})$.
Significant structures are excluded to a level of 10$^{-2}$ ($5 \cdot 10^{-2}$)
for the near-threshold settings, and 10$^{-4}$ ($10^{-2}$)
for the quasielastic settings
compared to the height of the neutron peak, respectively.
 
\section{Summary}
\label{sec:summary}
We have searched for
narrow structures in the missing-mass spectra of the electromagnetic
H(e,e$'\pi^{+}$)X and D(e,e$'$p)X reactions in kinematics 
corresponding to the hadronic beam experiments of Tatischeff {\it et
  al.}~\cite{tati} 
and Fil'kov {\it et al.}~\cite{filkov}. 
Our measurements exclude
narrow nucleon resonances below pion threshold  for the H(e,e$'\pi^{+}$)X
reaction 
down to the level of 10$^{-4}$
compared to the respective
transition to the neutron ground state.
For the D(e,e$'$p)X reaction,
narrow structures are excluded to a level of 10$^{-4}$ relative to the
neutron peak in the quasielastic 
and to 10$^{-2}$ in the near-threshold measurements.
On the basis of those results the various theoretical
attempts~\cite{konno,kobushkin,walcher} for an
explanation of the observed structures in Refs.~\cite{tati,filkov} might be
questioned. In order to draw a final conclusion on the existence of the narrow
resonances of Refs.~\cite{tati,filkov}, the hadronic experiments should be
repeated with the same accuracy as the present electromagnetic ones.

\section{Acknowledgement}
\label{sec:acknowledgement}
This work has been supported by the Deutsche Forschungsgemeinschaft
(SFB 443 and RI 242/15-2).


\begin{thebibliography}{99}
\bibitem{thomasweise}
        A.~Thomas and W.~Weise, 
        {\it The Structure of the Nucleon} (Wiley-VCH, Berlin, 2001).
\bibitem{tati} 
        B.~Tatischeff, J.~Yonnet, N.~Willis, M.~Boivin, M.P.~Comets,
        P.~Courtat, R.~Gacougnolle, Y.~Le~Bornec, E.~Loireleux, and
        F.~Reide, 
        Phys. Rev. Lett. {\bf{79}}, 601 (1997).

\bibitem{filkov}
        L.V. Fil'kov, V.L. Kashevarov, E.S. Konobeevski, M.V. Mordovskoy,
        S.I. Potashev, V.A. Simonov, V.M. Skorkin, and S.V. Zuev,
        Eur. Phys. J. {\textbf{A12}}, 369 (2001).

\bibitem{tati2} 
        B.~Tatischeff, J.~Yonnet, M.~Boivin, M.P.~Comets, P.~Courtat, 
        R.~Gacougnolle, Y.~Le~Bornec, E.~Loireleux, F.~Reide, and N.~Willis, 
        nucl-ex/0207003;
        B.~Tatischeff, nucl-ex/0207004.

\bibitem{mulders}
        P.J.~Mulders, A.T.~Aerts, and J.J.~de~Swart,
        Phys. Rev. {\bf{D21}}, 2653 (1980); 
        Phys. Rev. {\bf{D19}}, 2635 (1979); 
        Phys. Rev. Lett. {\bf{40}}, 1543 (1978).

\bibitem{konno}
        N.~Konno, 
        Nuov.~Cim. {\textbf{A111}}, 1393 (1998).

\bibitem{lvov}
        A.I.~L'vov and R.L.~Workman, 
        Phys. Rev. Lett. {\textbf 81}, 1346 (1998).

\bibitem{leon}
        V.~Olmos~de~Leon, F.~Wissmann, P.~Achenbach, J.~Ahrens, H.J.~Arends, 
        R.~Beck, P.D.~Harty, V.~Hejny, P.~Jennewein, M.~Kotulla,
        B.~Krusche, V.~Kuhr, R.~Leukel, J.C.~McGeorge, V.~Metag, R.~Novotny, 
        A.~Polonski, F.~Rambo, A.~Schmidt, M.~Schumacher, U.~Siodlaczek, 
        H.~Str\"oher, A.~Thomas, J.~Wei\ss, and M.~Wolf,
        Eur. Phys. J. {\textbf{A10}}, 207 (2001), and references therein.

\bibitem{kobushkin}
        A.P.~Kobushkin,
        nucl-th/9804069.

\bibitem{walcher}
        Th.~Walcher,
        hep-ph/0111279. %, submitted to Phys. Rev. Lett.

\bibitem{kolomeitsev}
        E.E.~Kolomeitsev and D.N.~Voskresensky,
        nucl-th/0207091.

\bibitem{jlab}
        X.~Jiang, R.~Gilman, R.~Ransome, P.~Markowitz, T.-H.~Chang,
        C.-C.~Chang, G.A.~Peterson, D.W.~Higinbotham, M.K.~Jones, N.~Liyanage,
        and J.~Mitchell, 
        Phys. Rev. {\bf C67}, 028201 (2003).
%        X.~Jiang, R.~Gilman, R.~Ransome, P.~Markowitz, T.-H.~Chang,
%        C.-C.~Chang, G.A.~Peterson, D.W.~Higinbotham, M.K.~Jones, N.~Liyanage,
%        J.~Mitchell, and B.~Wojtsekhowski, 
%        nucl-ex/0208008, submitted to Phys. Rev. {\bf{C}}.

\bibitem{tamii}
        A.~Tamii, K.~Hatanaka, M.~Hatano, D.~Hirooka, J.~Kamiya, H.~Kato,
        Y.~Maeda, T.~Saito, H.~Sakai, S.~Sakoda, K.~Sekiguchi, N.~Uchigashima,
        T.~Uesaka, T.~Wakasa, and K.~Yako, 
        Phys. Rev. {\bf C65}, 047001 (2002).

\bibitem{filkov2}
        L.V. Fil'kov, nucl-th/0208028.

\bibitem{beck}
        R.~Beck, S.N.~Cherepnya, L.V.~Fil'kov, V.L.~Kashevarov, M.~Rost, 
        and Th.~Walcher,  
        Proc. Int. Workshop on the Physics of Excited Nucleons, 
        Johannes Gutenberg-Universit\"at Mainz, Germany, March 7 - 10, 2001,
        nucl-th/0104070, World Scientific (Singapore), in press.

\bibitem{MAMI}   
        K.I.~Blomqvist, W.U.~Boeglin, R.~B\"{o}hm, M.~Distler,
        R.~Edelhoff, J.~Friedrich, R.~Geiges, P.~Jennewein, 
        M.~Kahrau, M.~Korn, H.~Kramer, K.W.~Krygier, V.~Kunde, 
        A.~Liesenfeld, H.~Merkel, K.~Merle, U.~M\"uller, 
        R.~Neuhausen, E.A.J.M.~Offermann, Th.~Pospischil,
        A.W.~Richter,  G.~Rosner,  P.~Sauer, St.~Schardt, 
        H.~Schmieden, A.~Wagner, Th.~Walcher, and St.~Wolf,
        Nucl. Instr. and Meth. {\bf A403}, 263 (1998).

\bibitem{prop98} 
        MAMI-Proposal A1/2-98 (1998).

\bibitem{maid}
        D.~Drechsel, O.~Hanstein, S.S.~Kamalov, and L.~Tiator,
        Nucl.\ Phys.\ {\bf A645}, 145 (1999).

\bibitem{gen} 
        M. Seimetz, Proc. 9$^{th}$ Int. Conference on the Structure of 
        Baryons, Jefferson Lab, Newport News, March 3 - 8, 2002,
        World Scientific (Singapore), in press.

\bibitem{arenhoevel}
        H.~Arenh\"ovel, W.~Leidemann, and E.L.~Tomusiak,
        Phys.\ Rev.\ {\bf C46}, 455 (1992).

\end{thebibliography}
\end{document}